\documentclass[aps, prd, twocolumn, lengthcheck, superscriptaddress, % showpacs,
nofootinbib]{revtex4-1}

\usepackage{epsfig}
\usepackage[usenames]{color}
\usepackage{graphicx}
\usepackage{amsmath}
\usepackage{epstopdf}

\newcommand\sect[1]{\emph{#1.}---}
\def\bi{\bibitem}

\def\la{\langle}\def\ra{\rangle}
\def\be{\begin{eqnarray}}\def\ee{\end{eqnarray}}
\def\lsim{\mathrel{\rlap{\lower3pt\hbox{\hskip1pt$\sim$}}
     \raise1pt\hbox{$<$}}} %less than or approx. symbol
\def\gsim{\mathrel{\rlap{\lower3pt\hbox{\hskip1pt$\sim$}}
     \raise1pt\hbox{$>$}}} %greater than or approx. symbol
\def\del{\partial}

\def\calL{\cal L}

\allowdisplaybreaks

\begin{document}

\title{How and How Much is ${g_A}$ {\it Fundamentally} Quenched in Nuclei?}

%\author{Yong-Liang Ma}
%\email{yongliangma@jlu.edu.cn}
%\affiliation{Center for Theoretical Physics and College of Physics, Jilin University, Changchun, 130012, China}

\author{Mannque Rho}
\email{mannque.rho@ipht.fr}
\affiliation{Universit\'e Paris-Saclay, CNRS, CEA, Institut de Physique Th\'eorique, 91191, Gif-sur-Yvette, France }

\date{\today}

\begin{abstract}

The superallowed Gamow-Teller transition in the doubly-magic-shell nucleus $^{100}$Sn and the high resolution spectral  shape analysis in the fourth-forbidden nonunique transition in $^{115}$In indicate as much as $\sim 40\%$ {\it fundamental} quenching in the axial-current coupling constant $g_A$ in nuclei.  This can be attributed to an effect of the trace anomaly in QCD ``emerging" in nuclear medium. If confirmed, this would signal a major revamping to do in nuclear interactions consistent with chiral-scale symmetry in nuclear medium and a big impact on $0\nu$ and $\nu\nu$ double $\beta$ decays for BSM. I present an argument that such a big anomaly-induced quenching is incompatible with how hidden scale symmetry manifests in nuclear medium,  A possible means to resolve this issue is discussed in terms of hidden scale symmetry permeating in baryonic matter from normal nuclear matter to massive compact-star matter.
\end{abstract}

\maketitle

\sect{Introduction}
In a recent Letter article~\cite{gA-MR}, it was agued that the long-standing ``puzzle" of the quenching of the axial coupling constant $g_A$ from the free-space value of $g_A=1.276$ to $g_A^\ast\simeq 1$ observed in Gamow-Teller transitions in light nuclei~\cite{gA-review} is resolved almost entirely by nuclear correlations. The quenching does not involve, if any,  a significant ``fundamental quenching"  of the axial vector constant $g_A$, but it is the {\it single-particle} $\sigma\tau$ operator that is quenched, that is, purely nuclear correlation effect.  There is absolutely nothing new in this result. This conclusion was  arrived at by many authors as listed e.g. in the reviews~\cite{gA-review}. To quote one example out of many, in 1977 Les Houches Lectures, Wilkinson concluded, based on his analysis, that with the shell-model wave-functions for light nuclei $A\leq 21$ that {\it include} ``full mixing,"   the ``effective" $g_{Ae}$ comes out to be~\cite{wilkinson}
\be
g_{Ae}/g_A=1+(2\pm 6)\%.
\ee
 In other words, there was nothing ``fundamental" in $g_A^\ast$ going down to 1 from $g_A=1.276$ listed in the Particle Physics Booklet. What was intriguingly mysterious was that $g_A^\ast$ was tantalizingly close to 1, resembling something fundamental like $g_V=1$ in CVC. Indeed the modern high-power many-body calculations such as  Quantum Monte Carlo calculations explain the $\beta$ decay rates for light nuclei $A < 10$ with the free-space $g_A$, modulo some small corrections coming from two-body exchange currents~\cite{wiringa}. 
 
 The situation in heavier nuclei was unclear~\cite{gA-review}.
 
In this note, I will describe what the solution to the puzzle could be,  what it implies how hidden scale symmetry manifests in nuclear dynamics and what fundamental information vis-\`a-vis with the quenched $g_A$ it could encode.

\sect{Resolving the puzzle}
What's new in the solution to the puzzle was  that it involved a  novel mechanism, so far totally unexplored,  based on a symmetry hidden in QCD that ``emerges" in strong nuclear correlations. It is argued that $g_A^\ast=1$ applies not only to light nuclei as observed~\cite{gA-review} but also to heavy nuclei as well as to dense matter: It is found to permeate from nuclear-matter density $n_0\simeq 0.16$ fm$^{-3}$ to high densities $n_{star}\sim 7 n_0$ relevant to compact-star physics, and beyond to what is referred to as ``dilaton-limit fixed point (DLFP)" $n_{\rm dlfp} \gsim 25 n_0$. 

The argument used in arriving at the  result in \cite{gA-MR} relied on nuclear effective field theory (EFT), coined  G$n$EFT, formulated to primarily address compact-star physics. It is formulated, however, to be valid for nuclear matter at  $\sim n_0$ for consistency with an accuracy comparable to that of the standard nuclear chiral EFT carried out to N$^p$LO, $p\sim 3$. Furthermore it is rendered applicable to compact-star physics~\cite{MR-review} by implementing the putative ``hadron-quark continuity (HQC)" conjectured to hold in QCD by a topology change  at $\sim 3 n_0$. The effective Lagrangian does not  involve phase transitions, hence no explicit quarks and gluons. Up to date, the results on compact-star properties fare surprisingly well, so far with no serious tension with the available observations~\cite{lattimer}. 

It is this G$n$EFT  that was applied to the $g^\ast_A$ problem in \cite{gA-MR}.  What's exploited there as a theoretical tool was the ``Fermi-liquid fixed-point (FLFP)" approximation in Landau Fermi-liquid theory~\cite{shankar}\footnote{The renormalization-group approach formulated for strongly-correlated electron system on the Fermi sea~\cite{shankar} was incorporated into chiral effective Lagrangian for nuclear interactions~\cite{FR}. It is identified as Landau-Migdal theory~\cite{migdal} since the pion plays a key role in nuclear physics.}  which corresponds to doing mean-field approximation in G$n$EFT. It turns out to be most closely mappable to a shell-model calculation of superallowed Gamow-Teller transitions in heavy doubly-magic-shell nuclei.  The heaviest such doubly-magic-shell nucleus known is $^{100}$Sn.   As in \cite{gA-MR}, I will focus on the Gamow-Teller transition from the ground state of $^{100}$Sn to the first excited $1^+$ state in $^{100}$In.  The key reasoning made in \cite{gA-MR} was that the FLFP calculation, which is valid at the large $\bar{N}$ limit \footnote{$\bar{N}\sim 1/k_F$ is defined as  $\bar{N}=k_F/(\Lambda_{\rm FS} -k_F)$ with $\Lambda_{\rm FS}$ standing for the cutoff on top of the Fermi surface.} can be identified with the ``Extreme Single Particle Model (ESPM)"  calculation of the superallowed Gamow-Teller beta decay of $^{100}$Sn. As far as I am aware, this equivalence can be justified neither in light nuclei nor in non-doubly-magic-shell heavy nuclei.

What's most noteworthy of the recent RIKEN measurement on this transition~\cite{RIKEN}, claimed to be improved from all previous measurements,  is that it deviates from what is expected if it agrees with the FLFP in Fermi-liquid theory
\be
g_A^\ast=g_A^{\rm L} \approx 1.\label{FLFPgA}
\ee
To match with the FLFP theory with $1/\bar{N}\to 0$, the Gamow-Teller strength ${\cal B}_{GT}$ should be
\be
{\cal B}^{\rm Landau}_{\rm GT}\approx 11.\label{Blandau}
\ee
The RIKEN experiment came out to be much lower,  $ {\cal B}_{\rm GT}^{\rm RIKEN}=4.4^{+0.9}_{-0.7}$.
Taking the central value, one sees that there is roughly 40\% quenching of $g_A$ compared with the FLFP value  (\ref{FLFPgA}). I will interpret this quenching as a possible ``fundamental renormalization" of $g_A$  due to the trace anomaly of QCD. This will be referred to as anomaly-induced quenching (acronymed $AIQ$) of $g_A$ expressed by $q_{\rm ssb} < 1$ (with $q_{\rm ssb}$ defined below in (\ref{qssb})). 
So far there has been no information on $\beta^\prime$ for $N_f \lsim 8$ in QCD, so this would be the first signal for it in hadronic physics coming from the nuclear physics side. 

In contrast to this RIKEN result, the data from the older GSi measurement~\cite{GSI} in which the daughter state is identified with a single-state,  
leading instead to ${\cal B}_{\rm GT}^{\rm GSI}\approx 10$, showed no indication of deviating from (\ref{FLFPgA}).

This discrepancy, if real, could present a serious issue with the $AIQ$. The RIKEN result in the $^{100}$Sn decay, if further confirmed, will raise a serious problem in nuclear physics and more importantly in both $2\nu$ and $0\nu$ $\beta\beta$ decay processes. Furthermore   most recent developments in experiments and theoretical analyses in the spectrum shape of multiply forbidden nonunique $\beta$ decays -- which involves  axial operators different from the superallowed GT -- in heavy nuclei indicate an equally important deviation of $q_{\rm ssb}$ from 1.  In comparing  the theoretical spectral shape  to the experimental spectrum in what is referred to as ``spectral-shape method (SSM)," a quenching of $g_A$ was noted to be $q\sim (0.6 - 0.7)$~\cite{SSM,spectral1,spectral2}. This amount of quenching  can be translated, as explained below, to $q_{\rm ssb}\approx  (0.6 - 0.7)$,  comparable to that observed in the RIKEN's $^{100}$Sn GT experiment. 

In what follows, I will argue  this quenching factor $q_{\rm ssb}\approx 0.6-0.7$   is at odds, in some case seriously, with what has been established in other nuclear weak processes.  

\sect{Fundamental renormalization by scale symmetry breaking}
What governs the $AIQ$ is how the scale symmetry breaking giving  the dilaton mass away from the infrared (IR) fixed point emerges in nuclear dynamics. Instead of the dilaton scalar $\sigma$ that transforms nonlinearly as the pion $\pi$ does in the chiral field $U=e^{i\pi/f_\pi}$, it is convenient to use the conformal dilaton compensator field $\chi=f_\chi e^{\sigma/f_\chi}$  which transforms linearly in scale transformation. As described in \cite{MR-review}, the argument relies on the ``genuine dilaton (GD)" approach to scale symmetry proposed by Crewther et al~\cite{GD,CT}\footnote{The ``genuine dilaton (GD)" is characterized by  the existence of an infrared (IR) fixed point at which both scale symmetry and chiral symmetry are spontaneously broken (``hidden") with the pions and the dilaton  as Nambu-Goldstone bosons excited but the matter fields, baryons and vector mesons, remain massive.}.  

At the classical level, the axial current coupling to the nucleon is scale-invariant. However there is an anomalous dimension contribution from the trace anomaly of QCD that enters nonperturbatively at the leading-order (LO) chiral-scale (CS) perturbation expansion~\cite{CT}. The axial current with the anomaly taken into account is of the form 
\be
J^{a\mu}_{ 5}=g_A q_{\rm ssb} \bar{\psi}\gamma^\mu\gamma_5 \frac{\tau^a}{2}\psi\label{AC}
\ee
where 
\be
q_{\rm ssb}= c_A+(1-c_A)(\frac{\chi}{f_\chi})^{\beta^\prime},\label{qssb}
\ee
$c_A$ is a constant that cannot be calculated perturbatively and $\beta^\prime$ is the derivative of the $\beta (\alpha_s)$ at the IR fixed point
\be
\beta^\prime|_{\alpha_s=\alpha_{\rm IR}} >0.
\ee 
In the matter-free space, the vacuum expectation value (VeV) is $\la\chi\ra=f_\chi$, so if one ignores the fluctuating dilaton field that one is justified to do for the axial current where the dilaton does not figure at the leading order,  one can set $q_{\rm ssb}=1$ and the current is scale symmetric: There is no $\beta^\prime$ effect. However in nuclear matter,  $\la\chi\ra^\ast=f^\ast_\chi\neq f_\chi$ brings in scale symmetry breaking, both explicit  and spontaneous, hence bringing in the $\beta^\prime$ dependence.  

Applied to nuclear matter, one then has the density-dependent factor $AIQ$
\be
 q^\ast_{\rm ssb} = c_A +(1-c_A)(\Phi^\ast)^{\beta^\prime}\label{aiq}
\ee
where
\be
\Phi^\ast= f_\chi^\ast/f_\chi \simeq  f_\pi^\ast/f_\pi\label{Phi}
\ee
with  $\ast$ standing for density dependence. It is reasonable to expect that $c_A$ would not strongly depend on density, so unless $c_A$ differs substantially from 1, one would expect that $\delta=1-q^\ast_{\rm ssb}$ would be $\ll 1$. I will now explain how this is not what seems to be the case. In fact it seems  the new (RIKEN)  result of $^{100}$Sn as well as the spectral shape results indicate that the $AIQ$ is 
\be
q^{^{100}\rm Sn}_{\rm ssb}\sim 0.6.\label{Sn}
\ee
This would imply that in nuclear medium, the ``fundamental" axial coupling constant is {\it not} $g_A =1.276$ as measured in the neutron $\beta$ decay but less than 1.

\sect{Hidden scale symmetry, nuclear correlations and fundamental quenching}
As mentioned, what's involved in the  SSM in heavy nuclei is drastically different from what's involved in  the superallowed Gamow-Teller (GT) transition in the doubly magic nucleus $^{100}$Sn. The difference is important.

To clarify what this means, it is worth recalling the strategy of mapping  the FLFP approximation in Fermi-liquid theory to the ESPM calculation in shell model of the  superallowed GT transition. The RIKEN quenching factor -- which comes out to be $q_{\rm RIKEN}\approx 0.5$  -- consists of two factors, the full nuclear correlation factor  $q_{\rm SNC}^{\rm Landau}\simeq 0.8$ derived in \cite{gA-MR} and  the $AIQ$ factor $q_{\rm ssb}\approx 0.6$ whereas the result of Hinke et al.~\cite{GSI}  is  $q_{\rm GSI}\approx  0.8$ implying $q_{\rm ssb}\approx 1$. The difference between the two may lie in how to experimentally {\it pinpoint} the daughter state in the ESPM. The argument based on Fermi-liquid theory in equating it  to the EPSM  is that the daughter state is entirely or at least dominantly populated by one shell-model configuration.   Now the problem is how to reliably identify the final-state  configuration in the measurement.  While this seemed feasible in \cite{GSI} which claimed that  the daughter state can be identified by the single $1^+$ state in $^{100}$In predicted to be populated by more than 95\% in an excitation energy of $\sim 3$ MeV -- which accounted for ${\cal B}_{\rm GT}^{\rm GSI}\approx 10$,  it is not clear whether this condition was met in \cite{RIKEN}. 

As for  the spectrum-shape analyses~\cite{SSM,spectral1,spectral2}, the axial-current operator is not directly governed by the hidden scale symmetry that plays the key role in the superallowed $^{100}$Sn transitions. In fact in accessing the $2\nu$ and $0\nu$ double beta decays the SSM is aimed to address, the transition operator that enters in the spectral shape is, in contrast to the $\sim 0$ momentum transfer in the allowed GT transitions,  highly forbidden,  involving up to $\sim 100$ MeV momentum transfers. Note also that  multifold-forbidden transitions in nuclei differ drastically from the unique first forbidden transition to which I will return below.

The systematic power counting on which the current successful standard nuclear chiral field theory is based works well in nuclei because the nuclear interactions largely involve energy-momentum scale of soft pions~\cite{weinberg}. Applied to the nuclear electroweak current~\cite{MR91}, this implies that the time component of the axial current $J^a_{5\mu}$ could receive an enhanced pion-exchange contribution~\cite{KDR}, but the space component which controls the GT transitions has a highly power-suppressed multi-body currents~\cite{KDR,tsp}. Unless the single-particle $\sigma\tau$ operator happens to be suppressed by kinematics or symmetry or others accidentally, the correction from the exchange currents should be strongly suppressed relative to the leading order (LO) ``impulse" term by N$^{z}$LO for $z\geq 3$. By the tenet of nuclear EFT, a suppression of this order in the power counting should be taken with extreme caution. Either a different counting scheme is adopted or such suppressed terms should be dropped.  If it turns out that a partial or incomplete sum of such corrections for $z=3$ is found to be ``important," then one cannot ignore terms with $z\geq 4$. .

Now the axial-current operator figuring in the SSM analyses~\cite{SSM,spectral1,spectral2} is of the ``impulse approximation" without multi-body current operators,  with, however,  configuration mixing (higher nuclear correlations) taken into account in various approximate ways. Given the multifold forbidden terms involved, it may very well be that the standard power counting in chiral EFT anchored on soft-ness kinematics makes no sense. So the corrections of the type applied to allowed and first-forbidden transitions could not even be formulated. Furthermore there is nothing like the ``Landau quenching factor" $q^{\rm Landau}_{\rm snc}$ that subsumes  the strongly correlated nuclear effect controlled by hidden scale symmetry~\cite{gA-MR}. It seems that what's done in \cite{SSM,spectral1,spectral2} is likely the best one can do at present.  It seems  fair to  assume that the theoretical treatment of SSM made in \cite{SSM,spectral1,spectral2} captures most  of the nuclear physics involved. Therefore the effective $g_A$ obtained, $g_A^\ast$, contains the $AIQ$.  This means that modulo the caveat that there is a tension with the decay rate as mentioned in the article,  the most recent high resolution spectral measurement~\cite{spectral2}  (of the four-fold forbidden decay of $^{115}$In) could be yielding the quenching factor
\be
q_{\rm ssb}^{{^{115}{\rm In}}} \approx 0.65-0.75\label{In}
\ee 
leading to the quenched $g^\ast_A\approx  0.83 - 0.96$.  This  is consistent with the $AIQ$ in the superallowed $^{100}$Sn GT transition (\ref{Sn}).

\sect{Is this $AIQ$ reasonable?}
The $\sim (30- 40)\%$ $AIQ$ (\ref{Sn}) and (\ref{In}) seems much too big. Can this be compatible with what's going on in nuclear physics? 

 As it stands, it immediately raises questions in various nuclear processes. 
 
 First of all, given that as defined (\ref{qssb}), the $AIQ$ is  a fundamental quantity at the level of a nuclear effective field theory, it should not depend appreciably on density except perhaps near phase changes. As noted above, there is no indication for it in the Monte Carlo calculation for  $A < 10$ and in the shell model for light nuclei $A < 21$. And there is no indication in nuclear processes where the constant $g_A$ could figure importantly. For instance through low-energy theorems such as the Goldberger-Treiman relation which should hold in nuclear medium, it figures, albeit indirectly,  without involving the EW currents and should affect nuclear processes where excitations of pionic quantum numbers -- such as the nuclear tensor forces -- are involved. Furthermore it enters in the calculation -- in nuclear EFT -- of the EoS of finite and infinite nuclei. So far no such effects have been observed.

One of the examples directly linked to the axial channel, in particular for the forbidden $\beta$ decays, is the first forbidden $0^\pm\leftrightarrow 0^\mp\ \Delta T=1$ transition. Unlike the many-fold forbidden transitions in the SSM, the operator for the first-forbidden transition is fairly well-defined and the transition is well measured  in heavy nuclei.  The first forbidden transition involves the time component of the axial current, $J^{a0}_{5}$, which for small momentum carried by the current gives a single-particle operator $\propto g^\ast_A {\sigma}\cdot {{p}}/m_N$ where ${p}$ is the nucleon momentum. It receives a strongly enhanced two-body exchange current operator, accurately calculated since it is dominated by a soft-pion exchange controlled by chiral symmetry.  The result obtained in experiments is usually expressed in terms of  $\epsilon_{\rm MEC}$ defined as the ratio of the matrix element of the axial-charge operator obtained from the data over $M_1$, the theoretical matrix element of the one-body axial-charge operator evaluated with the {\it unquenched} $g_A$. The theoretical expression for $\epsilon_{\rm MEC}$ with the soft-pion dominated two-body exchange-current term included but without $AIQ$ was first worked out in \cite{KR}. It is simple to incorporate the $AIQ$ factor in the result obtained in \cite{KR}. It takes the form (in G$n$EFT) 
%\be
%\epsilon_{\rm MEC}\equiv \frac{|M_1+M_2|}{|M_1|}
%\ee
%where . Now assuming that  $g_A$ is quenched by the $AIQ$ factor $q_{\rm ssb}$,  one obtains in G$n$EFT
\be
\epsilon^{q_{\rm ssb}}_{\rm MEC}=\frac{q_{\rm ssb}}{\Phi^\ast} \big(1+\frac{R}{\Phi^\ast} \big)\label{MEC}
\ee
where $R=M_2/q_{\rm ssb}M_1$ with $M_i$ with $i=1, 2$ standing, respectively, for the matrix element of the axial charge 1-body and 2-body operators without the $AIQ$ factor. $\epsilon^{q_{\rm ssb}=1}_{\rm MEC}$ (\ref{MEC}) was computed a long time ago for the Pb nucleus~\cite{KR}. Taking  $\Phi^\ast (n_0)\simeq 0.8$ extrapolated from measurements in deeply bound pionic atoms~\cite{yamazaki}, the prediction was
\be 
\epsilon_{\rm MEC}^{q_{\rm ssb}=1}\approx 2.0.\label{q=1}
\ee
With the quenching $q_{\rm ssb}\approx 0.6$, it gives the ratio
\be
\epsilon_{\rm MEC}^{q_{\rm ssb}=0.6}\approx 1.5.\label{q=0.6}
\ee
The experiment by Warburton in the Pb nuclei~\cite{warburton} gave 
\be
\epsilon_{\rm MEC}^{exp}=2.01\pm 0.05. 
\ee

The theoretical result (\ref{q=1}) has an uncertainty of about 10\% which covers the range of values in $\Phi$ extrapolated to nuclear matter density from the measured quantity in finite nuclei. The pion-exchange operator is dictated by chiral symmetry, so can be taken to be very accurate.  The wavefunctions used  may however be subject to improvement. Nonetheless the close agreement clearly favors the unquenched $g_A$, (\ref{q=1}). 

One may wonder how the $AIQ$ depends on density, that is, whether the $AIQ$, negligible in light nuclei (low density), increases at higher density where the significant $AIQ$ seems to manifest. In fact there has been some discussion on the possible effect of the $\beta^\prime$ in the range $1\lsim \beta^\prime\lsim 3$ on baryonic matter at high density involving chiral restoration~\cite{MR-omega}.  For QCD with the flavor $N_f\leq 3$, $\beta^\prime$ is unknown, therefore the $AIQ$ factor cannot be even roughly estimated. To have an idea of what $\beta^\prime$ can do, take $\beta^\prime <2$ considered for dense matter which may not be applicable to the present problem. One finds that  for this value of the $\beta^\prime$, the  $AIQ$ both for$^{100}$Sn decay  and $^{105}$In could be in a clear tension with Eq.~(\ref{aiq}).  What happens at high density is an open issue.

\sect{Effect of many-body currents in quenching}
In a recent Nature article~\cite{firstprinciple}, an argument was given that by means of changing the ``resolution scale" (or the cut-off scale of EFT), most of the quenching can be shifted from one-body  to two-body exchange-current (2BC) effects leading to  $q=0.75$ -- comparable to the Landau-Migdal Fermi-liquid fixed point value $q^L_{snc}=0.79$~\cite{gA-MR} --  in the $^{100}$Sn GT matrix element. Consequently no fundamental quenching was needed there.
%\footnote{It appears that the ``improved" RIKEN measurement was not taken into account in arriving at this result.}

There is a problem here however  concerning the 2BC. Unlike the axial-charge current for the first-forbidden transition mentioned above, the leading 2BC in GT transitions appears highly suppressed in chiral counting, appearing first at  N$^{\geq 3}$LO relative to the one-body $\sigma\tau$ operator~\cite{tsp}. This follows trivially from soft-pion theorems in the one-pion exchange graph. This was the other side of the coin --   no ``chiral filtering" protection  --  to the  side of  strongly enhanced soft-pion term in the $0^\pm\leftrightarrow 0^\mp$ transition, in what was coined as ``chiral filtering mechanism (CFM)." The tenet of nuclear EFT states that  strongly suppressed N$^n$LO terms, unless the lower order, say,  N$^{n-1}$LO (computed), term is accidentally small, cannot be trusted as in the Gamow-Teller transitions. If for instance N$^3$LO term is non-negligible, it makes sense only if N$^{\geq 4}$LO  terms are duly included,\footnote{An example for this  is given in \cite{wiringa} for $^8$Li, $^8$B and $^8$He beta decays. It unfortunately is not possible to check this matter at present since there are too many unknown constants to fix at higher order.} particularly in connection with the leading superallowed transition.  In \cite{firstprinciple}  the resolution scale is ``tweaked"  such that 2BC appearing at high chiral-order even started to ``dominate."  

An idea belonging to the same class of reasoning was made a long time ago when the entire quenching due to nuclear correlations was moved to $\Delta$-hole states by changing the resolution scale such that the $\Delta$-hole states in the relevant configuration space give the dominant contribution. Indeed the sum of $\Delta$-hole bubbles gave the $\Delta$-hole quenching $q_{\Delta}\approx 0.76$~\cite{delta}, just about what is obtained in \cite{firstprinciple} by the two-body currents. The caveat there was that the Landau-Migdal $g_0^\prime$ interaction parameter with universality assumption was unfounded, so the idea was dropped. 

\sect{Conclusion}
The effective $g_A^\ast$ for a quasiparticle undergoing superallowed Gamow-Teller transition  on top of the Fermi surface calculated in mean field with a simple nuclear chiral Lagrangian gave $g_A^\ast =g_A^L\simeq 1$~\cite{FR}. This was given a justification by the Fermi-liquid fixed point (FLFP) approach in a more sophisticated EFT,  G$n$EFT,  with heavy degrees of freedom encoded in hidden local symmetry and hidden scale symmetry~\cite{gA-MR}. What enters in $g_A^L$ is the product $\Phi^\ast \tilde{F}_1^\pi$ -- where $F_1^\pi$ is the pionic contribution to the Landau-Migdal interaction, both being Landau-Migdal fixed-point quantities. The product is remarkably insensitive to nuclear density in the vicinity of equilibrium nuclear matter,  so $g_A^L\simeq 1$ is suggested to apply to (heavy) finite nuclei as well as infinite nuclear matter. It is intriguing that $g_A^\ast\to 1$ even at the dilaton-limit fixed-point (DLFP) as discussed in \cite{MR-review} for compact stars.

In \cite{gA-MR}, it was suggested that $g_A^L$ could be fairly reliably  equated -- although a rigorous proof is lacking at present --  to the $g_A^\ast$ in the superallowed GT transition in $^{100}$Sn computed in the ``Extreme Single Particle (shell) Model."  It is this reasoning that led to the prediction of the Gamow-Teller strength for  the $^{100}$Sn transition ${\cal B}_{GT}$ to be  equal to  (\ref{Blandau}). While the GSI result was in agreement with this prediction with the daughter state identified  with a single ESPM configuration, the RIKEN result was clearly not. Supported by the SSM analysis of the spectral shape in strongly forbidden transitions, the RIKEN result indicated a $(30-40)\%$ anomaly-induced quenching of $g_A$. While such a quenching raises already a serious issue in nuclear dynamics (as in the unique first-forbidden transition), it will surely be a lot more important, involving the fourth power of $g_A$,  for the $0\nu$ beta decays relevant to going BSM. 

How does one go about resolving this issue?

It seems that given the possible theoretical simplicity of the superallowed transition in the doubly-magic-shell nuclei compared with the theoretically less well-controlled many-fold forbidden nonunique axial operators, revisiting the $^{100}$Sn decay -- and other doubly-magic shell nuclei if available -- is clearly in order. The close mapping between the Fermi-liquid fixed-point approach anchored on renormalization group flow and the ESPM structure in doubly-magic shell heavy nuclei offers the possibility  to identify experimentally, and to sharpen theoretically, what  the ESPM final state in $^{100}$In to which the transition takes place is. This poses a challenge to nuclear physics, both experimental and theoretical.

\end{document}